\documentstyle[twocolumn,prl,aps]{revtex}
\input epsf

\begin{document}

\draft 
\twocolumn[\hsize\textwidth\columnwidth\hsize\csname@twocolumnfalse\endcsname 

\title{String-like Clusters and  Cooperative Motion in  a Model
Glass-Forming Liquid}

\author{Claudio Donati$^1$, Jack F. Douglas$^{1,2}$, Walter Kob$^3$, 
Steven J. Plimpton$^4$, Peter H. Poole$^5$ and Sharon C. Glotzer$^{1,2}$}

\address{$^1$ Center for Theoretical and
Computational Materials Science, NIST, Gaithersburg, Maryland, USA 20899}

\address{$^2$ Polymers Division, NIST, Gaithersburg, Maryland, USA 20899}

\address{$^3$ Institut f\"ur Physik, Johannes Gutenberg Universit\"at,
Staudinger Weg 7, D-55099 Mainz, Germany}

\address{$^4$ Parallel Computational Sciences Department, 
Sandia National Laboratory, Albuquerque, NM 87185-1111}

\address{$^5$Department of Applied Mathematics,
University of Western Ontario, London, Ontario N6A~5B7, Canada}

\date{\today}
\maketitle

\begin{abstract}
A large-scale molecular dynamics simulation is performed on a
glass-forming Lennard-Jones mixture to determine the nature of
dynamical heterogeneities which arise in this model fragile liquid.
We observe that the most mobile particles exhibit a cooperative motion
in the form of string-like paths (``strings'') whose mean length and
radius of gyration increase as the liquid is cooled.  The length
distribution of the strings is found to be similar to that expected
for the equilibrium polymerization of linear polymer chains.
\end{abstract} 


\vskip2pc]
\narrowtext

Many glass-forming liquids at temperatures above their glass
transition exhibit unusual dynamical properties associated with
incipient glass formation\cite{science}.  In particular, relaxation
measurements (NMR, light scattering, dielectric, mechanical) on
glass-forming liquids typically give relaxation functions which are
non-exponential and these observations have been interpreted in terms
of ``dynamic heterogeneity'' in the liquid
structure~\cite{hetero,otp}.  Measurements of the translational and
rotational diffusion of tracer molecules in ``glassy liquids''
\cite{super} have shown the breakdown of continuum hydrodynamics and
these deviations have also been interpreted in terms of liquid
heterogeneity \cite{ediger}.  However, the typical optical clarity of
glassy liquids, as well as neutron scattering measurements over a wide
wavelength range, do not provide evidence for local density
fluctuations which could be readily associated with the
heterogeneities inferred from dynamic measurements \cite{neutron}.  It
therefore remains an open question whether glassy liquids contain
heterogeneities, and if so, what form they take.

The determination of the structure of glass-forming liquids requires
high resolution measurements in space and time.  Molecular dynamics
(MD) simulations of glassy liquids provide a powerful adjunct to
experimental studies, since the molecular structure and motions can be
examined in great detail.  However, this computational approach has
its own limitations since large, well equilibrated samples must be
followed for long enough times to explore the dynamics of glassy
liquids \cite{plimpton}. It is also not obvious how this microscopic
information should be best analyzed to reveal the nature of
heterogeneities in these liquids.

Previous MD simulations on glassy liquids have provided evidence for
their heterogeneity.  For example, dynamic heterogeneity has been
illustrated by averaging the properties of glass-forming liquids over
subregions to find that differences in these averages persisted for
significant times~\cite{thirumalai89}. Simulation has also shown that
the tracer diffusion of particles in glassy liquids deviates from the
Stokes-Einstein relation at low temperature~\cite{thirumalai93}, which
has been interpreted in previous measurements as implying
heterogeneity \cite{ediger}. Other MD studies have sought to examine
heterogeneity by tracking the motion of individual particles and have
revealed ample evidence for dynamic heterogeneity in the sense that
large dispersions in the particle mobilities in space and time were
apparent\cite{stupid}. However, no quantitative description of the
structural nature of heterogeneity in glass-forming liquids has
emerged from these studies. Here we return to this problem.

Extensive MD simulations of a glass-forming liquid in which the
molecules interact through a Lennard-Jones (LJ) potential were
performed to determine the existence and nature of heterogeneities in
this model liquid \cite{kdppg}.  The system studied in
Ref.~\cite{kdppg} was a three-dimensional binary mixture of 8000
particles having relative concentration of 80\% of type $A$ and 20\%
of type $B$ particles, where the size of the $A$ particles was 10\%
larger than the $B$ particles. The well depth $\epsilon$ and the
potential range $\sigma$ of the Lennard-Jones (L-J) potential,

\begin{equation}
V(r)=4\epsilon \Biggl[\biggl({{\sigma}\over{r}}\biggr)^{12} 
                     -\biggl({{\sigma}\over{r}}\biggr)^{6}\Biggr]
\label{lj}
\end{equation}

\noindent
and the relative particle concentration were chosen to suppress
crystallization and phase separation \cite{potential}.  Previous
studies of this system \cite{kob} have shown that it exhibits the
properties of a ``fragile'' \cite{A91} glass-forming liquid, and that
the relaxation of density fluctuations found for this model liquid
compares well with the predictions of mode-coupling theory.
Ref.~\cite{kdppg} reported evidence for dynamical heterogeneities in
this model glassy liquid, but their microscopic structure and dynamics
were not investigated in detail.  In this Report, we investigate the
simulated liquid described in Ref.~\cite{kdppg} in order to determine
the morphology of the heterogeneities that arise in this glassy liquid
\cite{macho}.

We report all quantities in dimensionless units involving the L-J
parameters of the $A$ particles \cite{units}.  The system was
equilibrated at 10 different temperatures $T$ in the range [0.451,
0.550]. For reference, the mode-coupling temperature $T_c$ was
estimated as $T_c=0.435$\cite{kob}.  The density varied from $1.09$
particles per unit volume at the highest $T$ to $1.19$ at the lowest
$T$ simulated.  Configuration histories for up to $4\times 10^6$ MD
steps following equilibration were stored for each run. Adopting 
argon values for the Lennard-Jones parameters of the larger spheres
\cite{rahman} gives an observational time of 26~ns.

\begin{figure}[tbc]
\hbox to\hsize{\epsfxsize=1.0\hsize\hfil\epsfbox{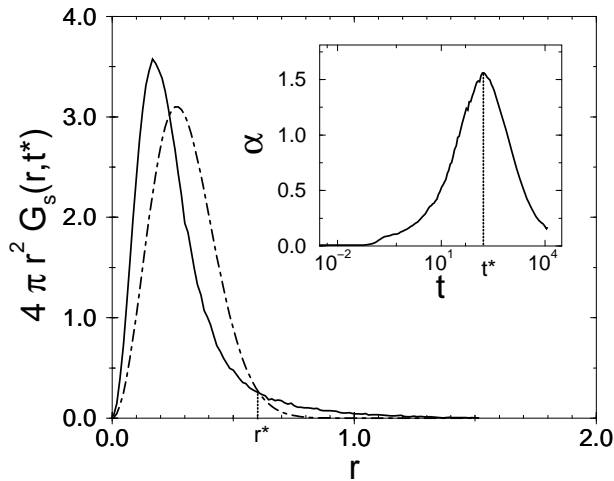}\hfil}
\caption{Self part of the van Hove correlation function $G_s(r,t)$
vs. $r$ at time $t^*$ when the non-Gaussian parameter $\alpha$ 
is a maximum (inset)~\protect\cite{kdppg}.  The dashed line
shows the expected curve if the distribution of particle displacements
was the Gaussian function $G_s^o(r,t)$ with $\langle r^2 \rangle$ equal to 
the simulation value
at $t^*$.  The mobile particles are defined as those particles for
which the particle displacement $r$ exceeds $r^*$.}
\label{fig:fig1}
\end{figure}

Ref.~\cite{kdppg} proposed a criterion for identifying particles of
enhanced mobility, and showed that the most mobile particles in the
liquid were spatially correlated.  In this approach, particle mobility
is assigned by comparing the self part of the van Hove correlation
function $G_s(r,t)$ with the Gaussian function $G_s^o(r,t)$

\begin{equation}
G_s^o(r,t)=\biggl({{3}\over{2\pi \langle r^2(t)\rangle}}\biggr)^{3/2}
\exp\biggl({{-3 r^2}\over{2 \langle r^2(t)\rangle}}\biggr)
\label{gauss}
\end{equation}

\noindent
describing the probability distribution of a Brownian particle with
displacement $r$ from the origin.  At long times, Brownian motion is
eventually established in a liquid, and hence $G_s(r,t)$ is well
approximated by $G_s^o(r,t)$ in the limit of large $t$.  $G_s^o(r,t)$
also describes a Brownian particle strongly localized by a harmonic
potential, where $\langle r^2\rangle$ is independent of $t$ .  At
short times the particles of a dense liquid are localized in the cage
created by surrounding particles, and hence $G_s(r,t)$ is also well
approximated by $G_s^o(r,t)$ in the limit of small $t$.  For
intermediate $t$, simulations of glassy liquids generally show
substantial deviations of $G_s(r,t)$ from $G_s^o(r,t)$
(Fig.~\ref{fig:fig1}).

\begin{figure}
\hbox to\hsize{\epsfxsize=0.9\hsize\hfil\epsfbox{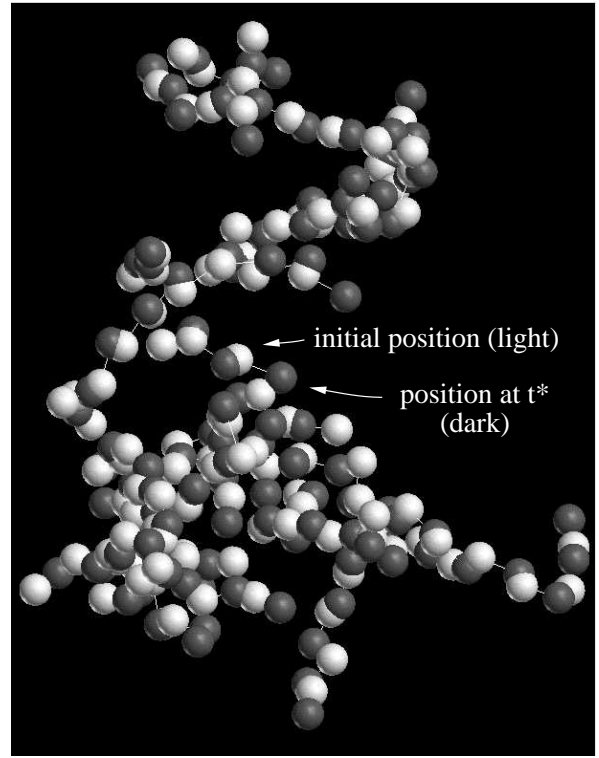}\hfil}
\bigskip
\caption{A snapshot of a large cluster of nearest-neighbor mobile
particles at $T=0.451$.  The lighter particles give the initial
particle positions, while the darker ones (shown connected to them via
line segments) designate the same particles at a later time
$t^*$. }
\label{fig:fig2a}
\end{figure}

\noindent
Deviation from the limiting Gaussian
$G_s^o(r,t)$ is conventionally measured by the ``non-Gaussian
parameter'' $\alpha(t)$ \cite{rahman}

\begin{equation}
\alpha(t)={{3 \langle r^4(t)\rangle}\over{5 \langle r^2(t)\rangle^2}} -1
\label{alpha}
\end{equation}

\noindent
which vanishes for $G_s^o(r,t)$.  Fig.~\ref{fig:fig1} shows $G_s(r,t)$
for the coldest simulation ($T=0.451$) at the time
$t^*$\cite{lifetime} where $\alpha(t)$ has a maximum (see
inset). Previous work has shown that dynamical heterogeneity is
strongly exhibited at this characteristic time\cite{stupid}. The long
tail of $G_s(r,t^*)$ relative to $G_s^o(r,t^*)$ shows that there exist
particles with enhanced mobility relative to ordinary Brownian motion
\cite{kdppg}. This point was also emphasized by Thirumalai and
Mountain\cite{thirumalai93}.  Following Ref.~\cite{kdppg}, we classify
a particle as ``mobile'' if its displacement $r$ at $t^*$ exceeds the
distance $r^*$ where $G_s(r,t)$ equals $G_s^o(r,t^*)$ (see
Fig.~\ref{fig:fig1}).

\begin{figure}
\hbox to\hsize{\epsfxsize=1.0\hsize\hfil\epsfbox{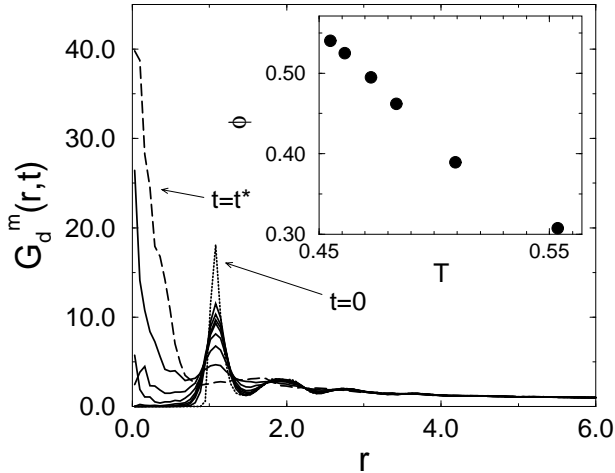}\hfil}
\caption{Distinct van Hove correlation function
$G_d^{m}(r,t)$ at various $t$. Inset: Fraction $\phi$ of mobile particles
that at $t^*$ are within a distance 0.6 from a position occupied by
another mobile particle at $t=0$.}
\label{fig:fig3}
\end{figure}

To analyze the geometrical configuration and dynamical motions of
these mobile particles, we define nearest-neighbor (nn) mobile
particles as those separated by a distance less than that of the first
minimum in the radial distribution function.  We can thereby identify
connected clusters of nn mobile particles.  A snapshot of a large
cluster of nn mobile particles in the coldest run is shown in
Fig.~\ref{fig:fig2a}.  The size distribution for these clusters is
broad, typically ranging from 1 to about 200 particles at the lowest
$T$. Note that many of the mobile particles are within large clusters
at low $T$, since the number of particles in the largest cluster is of
the same order of magnitude as the total number of mobile particles.
It should also be appreciated that this is a projection of a
three-dimensional structure, and consequently its diffuse, string-like
nature is somewhat obscured.  The average number $N_m$ of mobile
nearest neighbors of a mobile particle increases from $N_m=1.87$ at
the highest $T$ to $N_m=2.43$ at the lowest $T$, consistent with the
string-like appearance of the clusters (cf. Fig.~\ref{fig:fig2a}.
In comparison, the average coordination number of an atom in this
system is approximately 13.  The spatial extent of the cluster in
Fig.~\ref{fig:fig2a} is about 15 molecular units, which corresponds to
$\sim 3.3$~nm for orthoterphenyl, for which the hydrodynamic radius is
$0.21$~nm~\cite{otp}.

An examination of the particle motions indicates a correlated motion
of particles along quasi-one-dimensional paths.  To quantify the
nature of this correlated motion, we calculate the distinct van Hove
correlation function $G^{m}_d(r,t)$ for the mobile particles, which
describes the time-dependent density of particles in the vicinity of
an arbitrarily chosen test particle at $t=0$. The term ``distinct''
indicates that the test particle is excluded from the calculation of
the density, and at short times $G_d(r,t)$ is proportional to the
equilibrium radial distribution function $g(r)$. $G_d(r,t)$ is then a
time-dependent counterpart to $g(r)$, which provides information about
the degree of correlation in particle motions. For Brownian particles,
the peaks of $G_d(r,t)$, which are well developed at short times,
decay toward unity as particle positions decorrelate through Brownian
motion \cite{rahman}.  The depletion of particle density at small $t$
in $G_d(r,t)$ for $r<1$ gradually increases at later $t$ as other
particles move into the position previously occupied by the test
particle. If the position of the original test particle is occupied
with high probability by another mobile particle at later times,
$G^m_d(r,t)$ should develop a strong maximum about $r=0$.  This is
shown in Fig.~\ref{fig:fig3}, which presents $G_d^{m}(r,t)$ at
different $t$ for the lowest $T$.  At $t > 0$ the height of the nn
peak decreases, and a peak at $r=0$ develops and grows with time,
reaching its maximum value near $t \sim t^*$.  Note that no
significant peak exists for the $G_d(r,t)$ between mobile and other
particles.

\begin{figure}
\hbox to\hsize{\epsfxsize=1.0\hsize\hfil\epsfbox{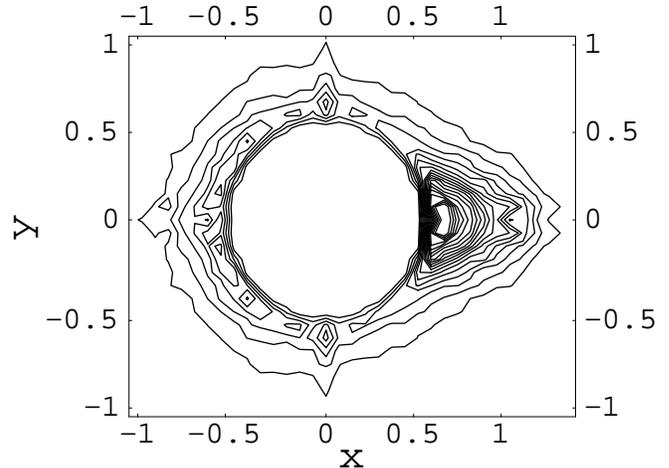}\hfil}
\caption{Contour plot of the probability distribution of the vector
$\delta \vec{r}_i(t^*)$ in the plane defined by $\delta
\vec{r}_i(t^*)$ and $\vec{r}_{ij}$. The vector $\vec{r}_{ij}$ defines
the positive $x$ axis.  The increment between any two contours is 5\%
of the total range. The distribution vanishes by definition in the
circular central region.}
\label{fig:fig4}
\end{figure}

The inset of Fig.~\ref{fig:fig3} shows the fraction $\phi$ of mobile
particles that are within a distance $\delta=0.6$ at $t^*$ from a
position occupied by another mobile particle at $t=0$~\cite{dist}.
The $T$-dependence of $\phi$, as well as that of the height of the
peak at $r=0$, indicates that the tendency for a mobile particle to
move toward a position previously occupied by another mobile particle
increases with decreasing $T$. Collective motion was recognized in
previous simulations of glass formation, but the effect was not
studied quantitatively. See e.g., \cite{strings}.

The particular string-like character of the collective particle motion
is demonstrated by examining the angular correlation between the
motion of two neighboring mobile particles.  Fig.~\ref{fig:fig4} shows
a contour plot of the probability distribution of the displacement
vector $\vec{\delta r}_i(t^*)=\vec{r}_i(t^*)-\vec{r}_i(0)$, projected
onto the plane defined by $\vec{\delta {r}}_i(t^*)$ and $\vec{r}_{ij}
\equiv \vec{r}_i(0)-\vec{r}_j(0)$, where $j$ denotes a mobile particle
within the nn shell of a mobile particle $i$. The vector
$\vec{r}_{ij}$ points in the direction of the positive $x$ axis. The
contours divide the range of values of the probability into 20
intervals so that the probability inside the innermost contour
centered on $(0.7,\ 0.0)$ is at least 20 times higher than the
probability outside the outermost contour. The distribution has been
calculated using all mobile particles that have at least one other
mobile particle in their nn shell.  If more than one mobile nn
particle exists in the proximity of a reference mobile particle, then
the averaging for the data in Fig.~\ref{fig:fig4} includes all
distinct pairs of nn mobile particles.  In the absence of any
correlation between $\vec{\delta r}_i(t^*)$ and $\vec{r}_{ij}$, the
contour plot should have a rotational symmetry about the origin.
Indeed, a contour plot calculated in a similar way between other (not
mobile) particles is rotationally symmetric.  Instead, the pronounced
asymmetry of the contour plot in Fig.~\ref{fig:fig4} shows that it is
much more probable for a mobile particle to move in the direction of
another mobile particle than in any other direction. It is this
dynamical correlation that appears to cause the string-like clustering
of mobile particles.

\begin{figure}
\hbox to\hsize{\epsfxsize=0.90\hsize\hfil\epsfbox{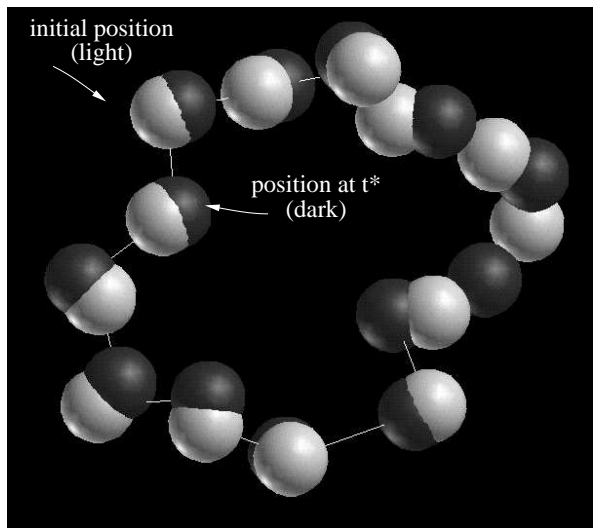}\hfil}
\bigskip
\caption{A large loop of 13 mobile particles exhibiting correlated
string-like motion at $T=0.451$.  Line segments connect identical
particles at successive times as in Fig.~2.}
\label{fig:fig2b}
\end{figure}

An animation of the cluster of nn mobile particles shown in
Fig.~\ref{fig:fig2a} reveals that not all of the mobile particles are
involved in string-like collective motion. These observations suggest
that we separately examine nn mobile particles which move collectively
in string-like paths. It is anticipated that these cooperatively
rearranging string-like clusters could have a large influence on
momentum \cite{egami,douglas} and mass transport \cite{ediger} in
glassy liquids, so we examine these structures in detail. The
potential significance of cooperative motion in glass-forming liquids
has long been appreciated \cite{crc}.

\begin{figure}
\hbox to\hsize{\epsfxsize=1.0\hsize\hfil\epsfbox{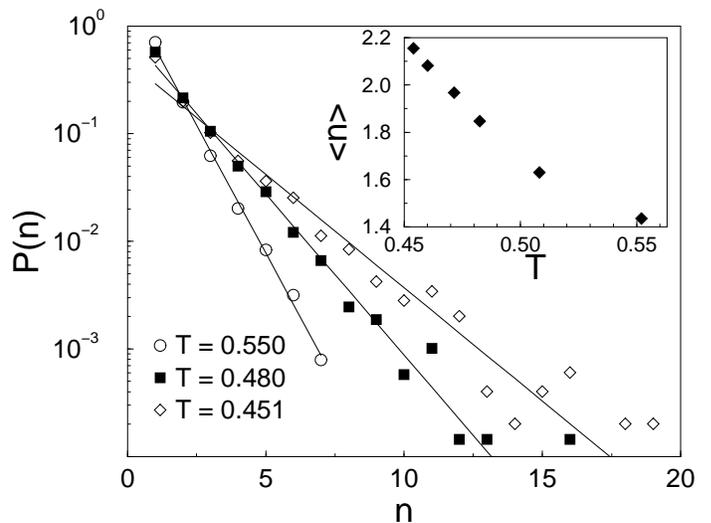}\hfil}
\caption{Probability distribution $P(n)$ of string lengths $n$ for 
various $T$. Inset: First moment $\langle n \rangle$ of $P(n)$ vs. $T$.}
\label{fig:fig5}
\end{figure}

We define ``strings'' by connecting two mobile particles $i$ and $j$
if min$[ |\vec{r}_i(t^*)-\vec{r}_j(0)|,|\vec{r}_i(0)-\vec{r}_j(t^*)|]
<0.6$.  This condition implies that one of the mobile particles has
moved, and a second mobile particle has occupied its
position. Fig.~\ref{fig:fig2b}\cite{movie} shows a representative
string which has a notable loop-like form. Most of the strings have
free ends, but loops occur with modest frequency. The strings have a
smaller average size than the nn mobile particle clusters. For example, 
the cluster in Fig.~\ref{fig:fig2a} consists of about 10
substituent strings.  The average length and size of the strings
increase as the liquid approaches the glass transition and their mass
$n$ distribution $P(n)$ is shown in Fig.~\ref{fig:fig5} for three
different $T$.  The semi-log plot reveals that $P(n)$ is approximately
exponential, i.e.  $P(n)\sim \exp(- n/\langle n \rangle)$, where the
average ``string length'' $\langle n\rangle$ increases with decreasing
$T$ (see inset).  For small $n$, the average radius of gyration $R_g$
of the strings appears to grow linearly with $\langle n \rangle$,
while for large $n$, an exponent near $1/2$ is observed, similar to
discrete random walk chains.  Observations similar to
Fig.~\ref{fig:fig5} have been found in studies of the equilibrium
polymerization of linear chain polymers \cite{eqpoly}.

By analogy with equilibrium polymerization \cite{eqpoly}, the fraction
of mobile particles involved in strings defines an ``order
parameter'' for describing the changing population of mobile
particles.  Since a particle exchange event involves at least two
particles, we count the fraction of mobile particles involved in
strings of length $n \ge 2$.  This fraction increases from $0.51$ for
the highest $T$ studied to $0.75$ for the lowest $T$, although the
total fraction of all particles that are mobile ($0.055 \pm 0.005$)
changes little over the $T$ range investigated. That is, an increasing
fraction of mobile particles participates in strings as the system is
cooled.

It should be noted that the occurrence of such dynamical
heterogeneities does not conflict with recent neutron scattering
observations \cite{neutron} which excluded the occurrence of {\it
compact} clustering in propylene glycol, a model organic glass former.
It is well known from previous neutron scattering studies on
equilibrium polymerization in, e.g., liquid sulphur \cite{sulphur}
that detecting such string-like structures by neutron scattering is
virtually impossible due to the small contrast between the chains and
the unpolymerized species. This difficulty should exist for other
polymerizing liquids as well, e.g. Te and Se \cite{Te}.

We can make some comparisons with other well-studied physical problems
where string-like structures have been predicted and observed.  The tendency
to form such structures, which involves a breaking of the local
rotational symmetry, is known to occur in, e.g., liquid crystals
\cite{yurke} and in many high energy physics contexts
\cite{highenergy,kibble}.  This type of spontaneous symmetry breaking
has been suggested previously for glassy liquids and glasses
\cite{rivier}, but previous investigations have not emphasized the
emergence of clusters which arise from correlated particle motion as
in the present study \cite{feynman}.

The observation of string-like heterogeneities in this cold, dense,
equilibrium Lennard-Jones liquid naturally leads to many other
questions about glassy liquids.  For example, do strings exist in
other model glass-formers?  Can they be observed in real liquids?
What is the nature of string-string interactions and how do these
interactions relate to the occurrence of the glass transition? Can the
characteristic properties of glasses at low $T$ (specific heat linear
in $T$, slowly varying thermal conductivity, etc.) be explained by the
present observation of string excitations? Are the excitations similar
in ``strong'' and ``fragile'' liquids \cite{douglas}? Does the
occurrence of physical aging in glasses reflect a structural evolution
of the string excitations? What effect does confinement have on the
dynamics of the strings and how does this relate to material
properties?  The study of the geometrical form of dynamical
heterogeneities in glassy liquids promises to lead to an improved
understanding of many aspects of glass formation.

We acknowledge useful conversations with Ray Mountain of NIST. 
PHP acknowledges the support of NSERC. WK is partially supported by 
Deutsche Forschungsgemeinschaft under SFB~262.


\begin{references}
\bibitem{potential} The L-J parameters $\epsilon_{\alpha\beta}$ and
$\sigma_{\alpha \beta}$ for $\alpha,\beta \in \{A,B\}$ are given by:
$\epsilon_{AA}=1.0$, $\sigma_{AA}=1.0$, $\epsilon_{AB}=1.5$,
$\sigma_{AB}=0.8$, $\epsilon_{BB}=0.5$, and $\sigma_{BB}=0.88$. Note
the presence of the substantial A-B interparticle interaction which
counterbalances the tendency for the size ratio
$\sigma_{AA}/\sigma_{BB}$ to cause phase separation.
\bibitem{macho} The simulations discussed here were run on the
massively parallel nCUBE 2 and Intel Paragon supercomputers at Sandia
National Laboratories.  In aggregate the equilibration and
statistics-gathering runs required approximately 300 hours of CPU time
on 256 Paragon nodes (i860 processors) and generated about 10 Gbytes of
data.
\bibitem{units} The length is defined in units of $\sigma_{AA}$,
temperature $T$ in units of $\epsilon_{AA}/k_B$, where $k_B$ is the
Boltzmann's constant, and time $t$ in units of
$\sqrt{\sigma_{AA}^2m/\epsilon_{AA}}$. The mass $m$ of both types of
particles are taken to be equal.
\bibitem{lifetime} The time $t^*$ corresponds to the end of the
transient plateau in $\langle r^2(t) \rangle$ where the particle motion crosses over from
a localized to asymptotic diffusive behavior.  This time was shown to
be related to the cluster lifetime in Ref.~\cite{kdppg}.
\bibitem{dist} 
The qualitative results are insensitive to the choice
of $\delta$, provided that $\delta$ is chosen smaller than $\sigma_{AA}$.
\end{references}
\end{document}